\documentclass[twocolumn,reprint, superscriptaddress,amsmath,amssymb,aps]{revtex4-2}
\usepackage[utf8]{inputenc}
\usepackage{graphicx}
\usepackage{dcolumn}
 
\usepackage{bm}
\usepackage{dirtytalk}
\usepackage{siunitx}
\usepackage{amssymb}
\usepackage{hyperref}
\usepackage{subfiles}
\usepackage{braket}

\begin{document}

\title{Tunable, phase-locked hard X-ray pulse sequences generated by a free-electron laser}

\author{Wenxiang~Hu}
\altaffiliation{These authors contributed equally to this work}
\affiliation{PSI Center for Photon Science, Paul Scherrer Institute, 5232 Villigen PSI, Switzerland}
\affiliation{Laboratory for Solid State Physics and Quantum Center, ETH Zurich, 8093 Zurich, Switzerland}

\author{Chi~Hyun~Shim}
\altaffiliation{These authors contributed equally to this work}

\author{Gyujin~Kim}
\author{Seongyeol~Kim}
\author{Seong-Hoon~Kwon}
\author{Chang-Ki~Min}
\author{Kook-Jin~Moon}
\author{Donghyun~Na}
\author{Young~Jin~Suh}
\author{Chang-Kyu~Sung}
\author{Haeryong~Yang}
\author{Hoon~Heo}
\author{Heung-Sik~Kang}
\affiliation{Pohang Accelerator Laboratory, POSTECH, Pohang, Gyeongbuk 37673, Republic of Korea}

\author{Inhyuk~Nam}
\email{ihnam@unist.ac.kr}
\affiliation{Pohang Accelerator Laboratory, POSTECH, Pohang, Gyeongbuk 37673, Republic of Korea}
\affiliation{Ulsan National Institute of Science and Technology, Ulsan, 44919, Republic of Korea}

\author{Eduard~Prat}
\affiliation{PSI Center for Accelerator Science and Engineering, Paul Scherrer Institute, 5232 Villigen PSI, Switzerland}

\author{Simon~Gerber}
\affiliation{PSI Center for Photon Science, Paul Scherrer Institute, 5232 Villigen PSI, Switzerland}

\author{Sven~Reiche}
\affiliation{PSI Center for Accelerator Science and Engineering, Paul Scherrer Institute, 5232 Villigen PSI, Switzerland}

\author{Gabriel~Aeppli}
\email{gabriel.aeppli@psi.ch}
\affiliation{PSI Center for Photon Science, Paul Scherrer Institute, 5232 Villigen PSI, Switzerland}
\affiliation{Laboratory for Solid State Physics and Quantum Center, ETH Zurich, 8093 Zurich, Switzerland}
\affiliation{Institute of Physics, EPF Lausanne, 1015 Lausanne, Switzerland}

\author{Myunghoon~Cho}
\email{mh0309@postech.ac.kr}
\affiliation{Pohang Accelerator Laboratory, POSTECH, Pohang, Gyeongbuk 37673, Republic of Korea}

\author{Philipp~Dijkstal}
\email{philipp.dijkstal@psi.ch}
\affiliation{PSI Center for Accelerator Science and Engineering, Paul Scherrer Institute, 5232 Villigen PSI, Switzerland}

\date{\today}


\begin{abstract}
\textbf{The ability to arbitrarily dial in amplitudes and phases enables the fundamental quantum state operations pioneered for microwaves and then infrared and visible wavelengths during the second half of the last century~\cite{Kawashima:1995,Krausz:2009,Cundiff:2010}.
Self-seeded X-ray free-electron lasers (FELs) routinely generate coherent, high-brightness, and ultrafast pulses for a wide range of experiments~\cite{Geloni:2016}, but have so far not achieved a comparable level of amplitude and phase control.
Here we report the first tunable phase-locked, ultrafast hard X-ray (PHLUX) pulses by implementing a recently proposed method~\cite{Reiche:2022}: A~fresh-bunch self-seeded FEL~\cite{Emma:2017}, driven by an electron beam that was shaped with a slotted foil~\cite{Emma:2004} and a corrugated wakefield structure~\cite{Emma:2014}, generates coherent radiation that is intensity-modulated on the femtosecond time scale.
We measure phase-locked (to within a shot-to-shot phase jitter corresponding to 0.1 attoseconds) pulse triplets with a photon energy of 9.7~keV, a pulse energy of several tens of microjoules, a freely tunable relative phase, and a pulse delay tunability between 4.5 and 11.9~fs.
Such pulse sequences are suitable for a wide range of applications, including coherent spectroscopy, and have amplitudes sufficient to enable hard X-ray quantum optics experiments.
More generally, these results represent an important step towards a hard X-ray arbitrary waveform generator.}
\end{abstract}


\maketitle

Phase-locked photon pulses underpin science and technology ranging from magnetic resonance imaging to quantum information~\cite{Ippen:1972,Becker:1972,Kawashima:1995,Krausz:2009,Cundiff:2010,Keller:2022}.
Figures 1a and 1b illustrate a sequence of coherent pulses in the time and frequency domains, respectively.
The high-harmonic generation (HHG) technique for optical lasers~\cite{Macklin:1993} provides access to such pulses down to extreme ultraviolet (XUV) and soft X-ray wavelengths~\cite{Pupeza:2021}.
However, HHG is limited in wavelength and pulse energy, primarily due to lower conversion efficiency at higher harmonics.
Phase-stable hard X-ray pulses were recently generated using atomic superfluorescence, but with a pulse energy far below \SI{1}{\uJ} and limited tunability~\cite{Zhang:2022}.
Tunable phase-locked hard X-ray pulses will present new opportunities for a wide range of experiments already used for longer photon wavelengths, including Ramsey spectroscopy~\cite{Ramsey:1950,Morgenweg:2014,Wituschek:2019}, time-domain ptychography~\cite{Spangenberg:2015}, cross-correlation measurements with infrared dressing field~\cite{Paul:2001,Maroju:2020,Maroju:2023}, and coherent control of quantum states~\cite{Greenland:2010,Nandi:2022,Nandi:2024}.
Additionally, resonant pulses with tunable delay and intensity could mitigate the radiation damage that affects X-ray diffraction and scattering experiments~\cite{Stohr:2015,Wu:2016,Chen:2018,Reiche:2022}, thereby offering new possibilities especially for the life sciences.

Free-electron lasers (FELs), driven by electron beams upon interaction with the magnetic fields of an undulator beamline~\cite{Pellegrini:2016,Seddon:2017,Huang:2021}, are sources of high-brilliance radiation with excellent wavelength tunability from the THz domain to the hard X-ray regime. Tunable phase-locked FEL pulses have so far been demonstrated in the THz~\cite{Greenland:2010}, XUV~\cite{Gauthier:2016,Wituschek:2019} and soft X-ray~\cite{Robles:2025} regimes.
External seed lasers are proving to be valuable for phase-locked FEL pulses~\cite{Thompson:2008, Prat:2023, Prat:2024}, but this approach is limited to generating pulse trains with spacing proportional to the seed laser wavelength, typically corresponding to a few femtoseconds. 

Self-seeding is a well established technique to achieve longitudinally coherent X-ray FEL radiation~\cite{Feldhaus:1997, Geloni:2011, Amann:2012, Ratner:2015, Geloni:2016, Emma:2017, Liu:2023}.
Therefore, phase-locked ultrafast X-ray (PHLUX) pulses could in principle be achieved with split-and-delay stages that produce replicas of self-seeded FEL pulses.
This technique is feasible in the  optical and THz regimes~\cite{Greenland:2010}, but not the hard X-ray regime due to the overly tight tolerances required for sub-wavelength (in this case of order 1 angstrom) path length stability~\cite{Reiche:2022}.
Recently, we proposed a more flexible method to generate PHLUX at a self-seeded FEL~\cite{Reiche:2022}, where instead of splitting the photon beam after the undulators, we split the electron beam before the undulators, which then generates separated and coherent X-ray pulses.
Here we describe the implementation and experimental demonstration of PHLUX, achieved in this way with a hundred-fold mechanical advantage over photon beam splitting. 

\section*{Experiment}

\begin{figure*}
  \centering
    \includegraphics[width=0.85\linewidth, trim=0cm 0cm 0cm 0cm, clip=true]{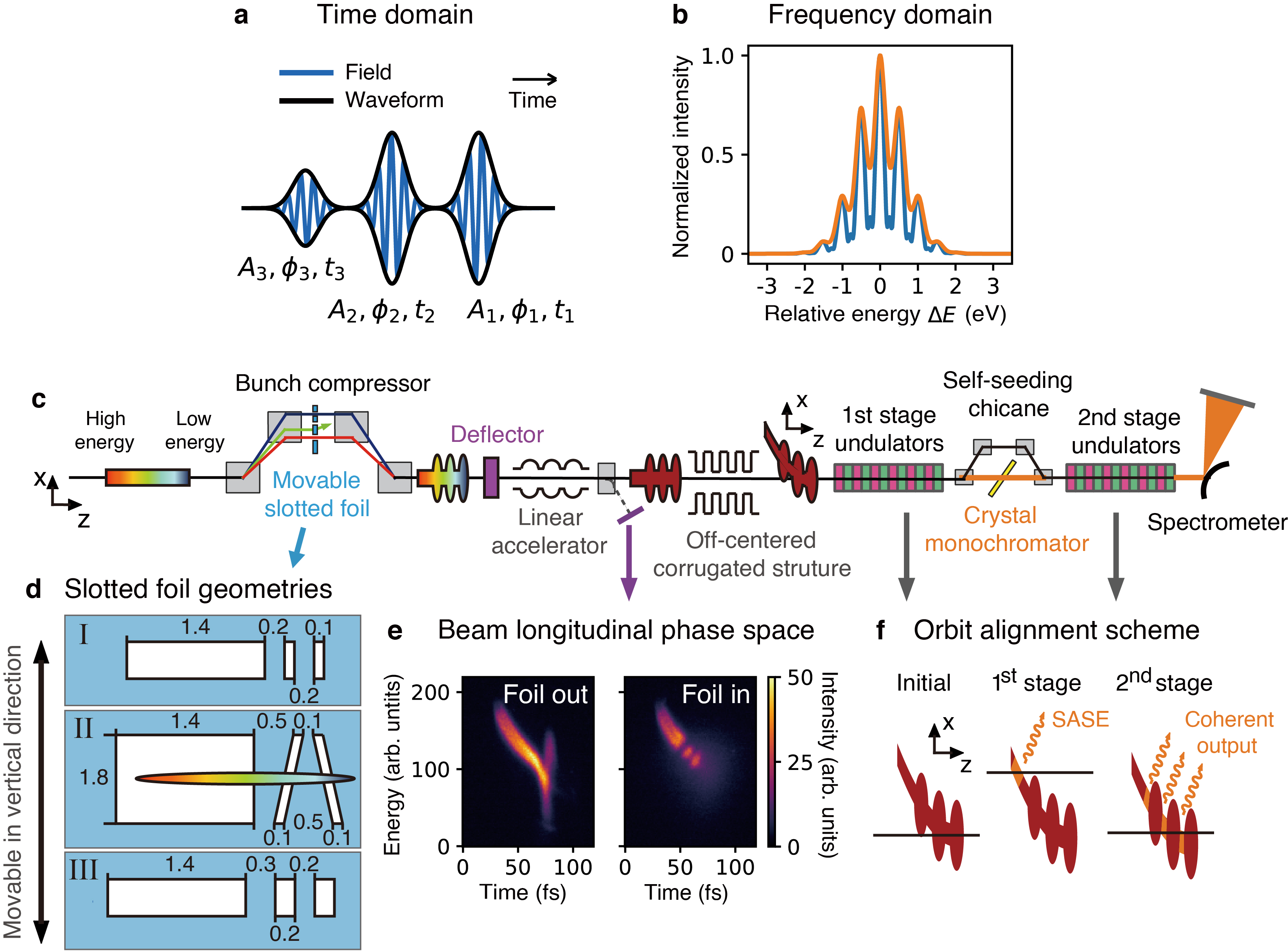}
    \caption{%
    \textbf{PHLUX generation at PAL-XFEL}.
    \textbf{a} Schematic PHLUX pulses in the time domain, characterized by individual pulse amplitudes~$A$, phases~$\phi$ and envelope peak arrival times~$t$. 
    \textbf{b} Schematic PHLUX pulses in the frequency domain (blue).
    The orange line shows the expected measured spectrum after considering the spectrometer resolution.
    \textbf{c} Accelerator layout for PHLUX generation.
    The undulator section is equipped for self-seeding with a delaying chicane and a monochromator.
    \textbf{d}~The final bunch compressor contains a movable slotted foil to ``spoil" the transverse emittance and energy spread of selected longitudinal parts of the beam. The time delay between electrons that pass different slots is proportional to the slot distance (given in units of millimeters). The available foil geometries I, II, and III contain a wide slit for SASE seed generation from the bunch tail, and two small slits for coherent lasing from the bunch head.
    \textbf{e} A radiofrequency deflector after the bunch compressor can measure the electron beam longitudinal phase space.
    The measurement without slotted foil (left) shows a uniform transverse beam size around the bunch center.
    After inserting the foil with geometry I (right), we measure three separate low-density regions owing to an increased transverse beam size.
    \textbf{f} Following further acceleration, the beam travels through a corrugated structure which imposes a beam tilt~\cite{Emma:2017}.
    The orbit control system then aligns the bunch tail to the ideal orbit in the first undulator stage and the bunch head in the second stage~\cite{Dijkstal:2020} for generation of SASE and coherent output, respectively.
    }
    \label{fig_1}
\end{figure*}

We apply the techniques of self-seeding, fresh-slice lasing, and slotted foil shaping at the Pohang Accelerator Laboratory X-ray FEL (PAL-XFEL)~\cite{Kang:2017} to realize hard X-ray PHLUX pulses.
Figure~\ref{fig_1}c shows the facility layout.
After preparation which we discuss in more detail below, the electron beam enters the undulators which are split into two stages.
The first stage operates in self-amplified spontaneous emission (SASE) mode and produces radiation with limited longitudinal coherence and a relatively large bandwidth.
A filter (here a crystal monochromator) selects a narrow band from the SASE spectrum, and the second FEL stage amplifies the transmitted, longitudinally coherent light pulse.
Fresh bunch self-seeding~\cite{Emma:2017} is a variant where distinct longitudinal subsets (slices) of the electron bunch emit radiation in the two stages.
Hence, the quality of the slices emitting coherent radiation in the second stage is not compromised by the SASE seed pulse emission, leading to higher brightness.
Furthermore, the pulse duration of the coherent radiation can be chosen independently from the SASE seed pulse duration.
Selection of such lasing slices is achieved through a correlation between transverse and longitudinal particle coordinates (beam tilt) and precise orbit control within the undulator beamline (Fig.~\ref{fig_1}f).
Slices not aligned to the ideal axis perform betatron orbit oscillations, which prevent them from amplifying radiation.
Additionally, a slotted foil placed in the bunch compressor irreversibly degrades the beam quality of selected slices such that they cannot emit FEL radiation~\cite{Emma:2004} (Fig.~\ref{fig_1}d,e).
The foil geometry can be manufactured with micron precision and controls the envelope of the coherent FEL radiation on a femtosecond time scale~\cite{Emma:2004, Reiche:2022}.
A temporal FEL intensity modulation with spacing $\Delta t$ corresponds to an interference spectrum with line splitting $\Delta E=h/\Delta t$ (where $h$~is the Planck constant).
Since the central photon energy of the seed is determined by the crystal monochromator, a stable interference pattern is expected.
The wavelength tunability of the monochromator enables also highly sensitive relative phase tunability: A change of central photon energy by $\Delta E$ is equivalent to a radiation phase change of 2$\pi$ among pulses delayed by $\Delta t$,
which is also reflected in an interference pattern shift by $\Delta E$.

\begin{figure*}
  \centering
    \includegraphics[width=\linewidth, trim=0cm 0cm 0cm 0cm, clip=true]{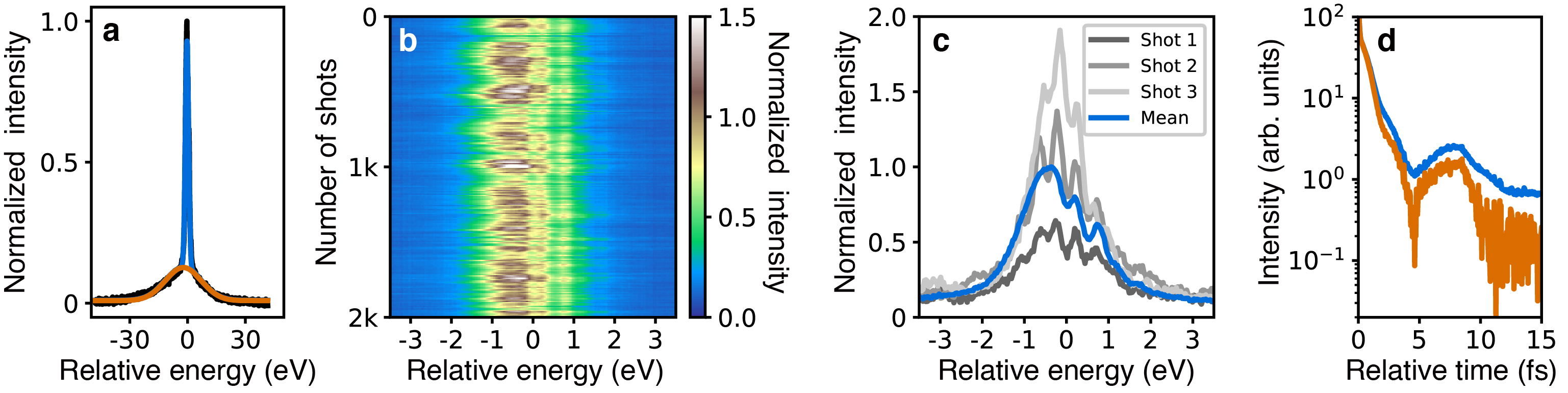}
    \caption{%
    \textbf{Spectral analysis of phase-locked pulses}.
    \textbf{a} The black line shows the average of 2,000 consecutive spectra acquired at a central photon energy of \SI{9.7}{keV}.
    The blue line depicts a fit of the spectrum using a sum of two Gaussian functions, while the orange line visualizes the Gaussian-shaped SASE background.
    \textbf{b} Coherent part of the 2,000 acquired spectra.
    \textbf{c} Three representative single-shot spectra (gray) and the average spectrum (blue).
    \textbf{d} We show the impact of the temporal FEL pulse modulation on the Fourier-transformed spectra in two different ways: the average amplitude of the single-shot (blue line) and the amplitude of the averaged single-shot Fourier transforms (orange line).
    }
    \label{fig_2}
\end{figure*}

\section*{Results}
We set up the fresh bunch self-seeding mode for a photon energy of $\SI{9.7}{keV}$ utilizing the nonlinear beam tilt generated by a corrugated wakefield structure~\cite{Emma:2017}, and then insert the slotted foil to generate PHLUX pulses.
We study the tunability using different foil geometries and self-seeding settings.
For FEL pulse characterization we use a single-shot spectrometer with a resolution of~\SI{0.26}{eV}~\cite{Kim:2025}.

As the transverse divergence of electrons hitting the slotted foil is greatly increased, they do not contribute to the FEL pulse generated in downstream undulators.
In the experiments, we use three slits to create three longitudinally separated regions with unspoiled electrons.
Electrons from the wide tail slit (\SI{1.4}{mm}) generate the SASE seed pulse.
The two head slits are narrower (\SI{0.1}{mm}), suitable for the generation of two short pulses in the second stage of the self-seeded FEL.
Due to the small separation between the tail and central slits, the orbit misalignment caused by the beam tilt does not entirely prevent a third coherent pulse from being emitted by electrons passing through the tail slit.
Consequently, our setup generates a coherent FEL pulse triplet (Fig.~\ref{fig_1}f).

To demonstrate PHLUX pulses, we generate radiation with an equal time separation between neighbouring pulses, resulting in a simplified spectral analysis.
Therefore, we use foil geometry~I (Fig.~\ref{fig_1}d) with equidistant slits.
For this first demonstration, we achieved an average total pulse energy of \SI{90}{\uJ} with a root mean square (rms) jitter of \SI{27}{\uJ} (\hyperref[supp]{Supplementary Information}).
Figure~\ref{fig_2}a shows the averaged spectrum of 2,000 consecutively acquired single-shot spectra, which includes a broadband SASE background typical for self-seeded FELs.
We fit the averaged spectrum with two Gaussian functions to obtain the ratio of the coherent signal intensity to the total pulse intensity, which is 41\% in this case.
Across different PHLUX measurements, the coherent ratio ranges from 40\% to 50\%.
Figure~\ref{fig_2}b shows these single-shot spectra with the plot range centered on the coherent part.
The vast majority of spectra contain a frequency modulation across the coherent part, which indicates a temporal amplitude modulation.
Due to relative phase stability, the average spectrum also contains interference lines (Fig.~\ref{fig_2}c).
To obtain the delay between neighbouring pulses we analyze the Fourier transforms of all spectra.
Both averages shown in Fig.~\ref{fig_2}d indicate a PHLUX pulse delay of $\SI{\approx8}{fs}$, corresponding to a frequency modulation period of \SI{0.5}{eV} \mbox{(\hyperref[Methods:correlation]{Methods})}.
We estimate the first-order temporal correlation function (\hyperref[Methods:correlation]{Methods}) at \SI{8}{fs} to be above 0.6.
Moreover, we deduce a delay of $\SI{\approx16}{fs}$ between the head and tail pulses and, thus, an additional spectrum modulation with \SI{0.25}{eV} period, which cannot be resolved by the available single-shot spectrometer.

\begin{figure*}
  \centering
    \includegraphics[width=0.8\linewidth, trim=0cm 0cm 0cm 0cm, clip=true]{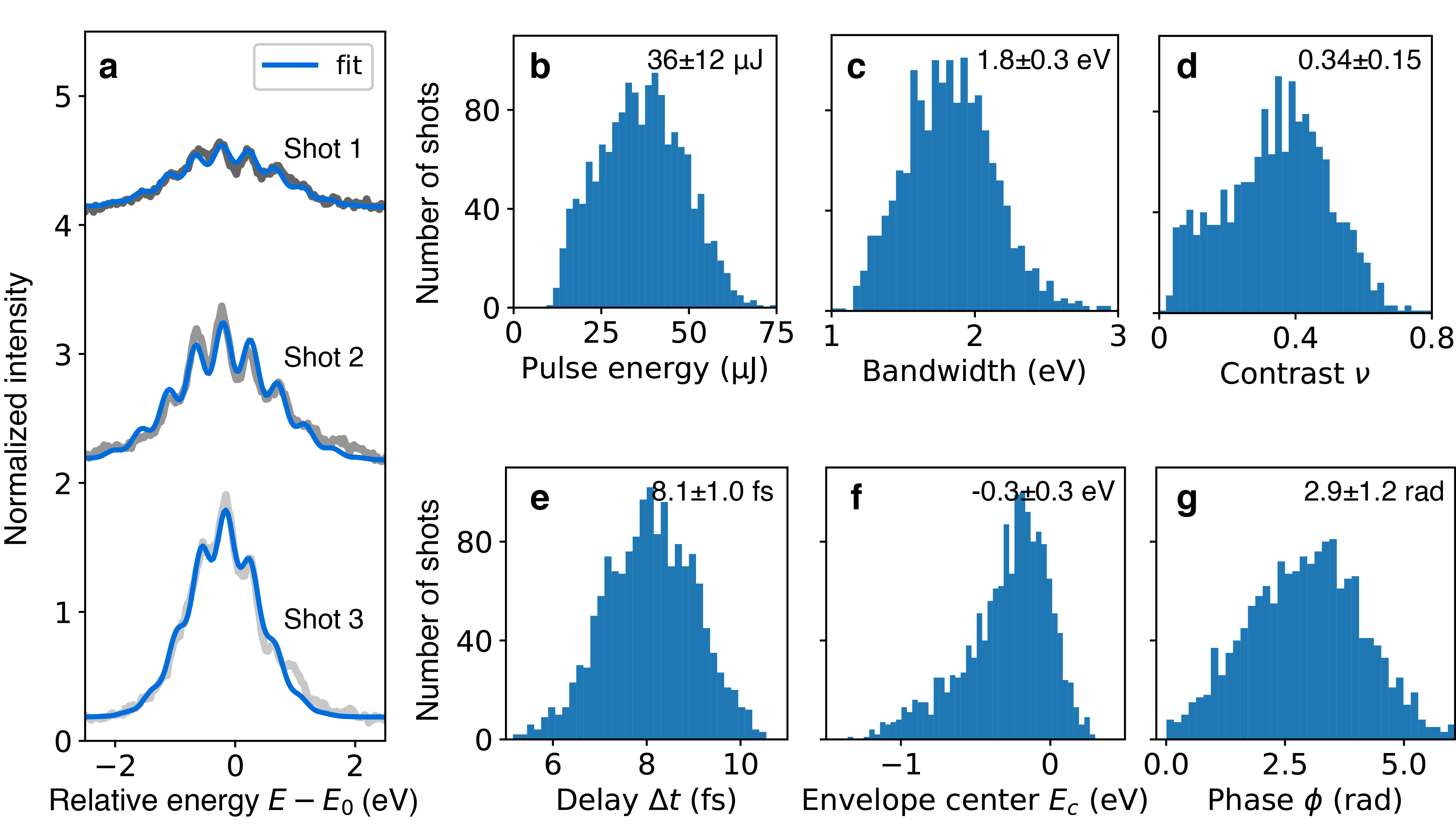}
    \caption{%
    \textbf{Single-shot spectra fits}.
    \textbf{a} Example single-shot spectra from Fig.~\ref{fig_2}c (gray) with corresponding fits to Eq.~(\ref{eq:fit}) (blue).
    \textbf{b-g} Histograms of the fitted coherent pulse energy (\textbf{b}), spectral FWHM bandwidth (\textbf{c}), contrast $\nu$ (\textbf{d}), pulse delay $\Delta t$ (\textbf{e}), envelope center position $E_{c}$ (\textbf{f}), and phase of the interference pattern $\phi$ (\textbf{g}).
    Numerical values indicate the average and the statistical variation.
    }
    \label{fig_3}
\end{figure*}

Knowing that three equally spaced coherent pulses are generated, we infer further time-domain information from the coherent part of the single-shot spectra.
Because the interference between head and tail pulses is not resolved, the spectral interference pattern (phase and contrast) corresponds to the beating between the middle and the sum of the other pulses.
We use a spectrum fit function that consists of a constant background (bg), a Gaussian envelope, and a sinusoidal amplitude modulation:
\begin{equation}
\label{eq:fit}
\begin{split}
    P(E)=&\mathrm{bg}+P_\mathrm{max}e^{-\frac{(E-E_{\mathrm{c}})^{2}}{2\sigma_{E}^{2}}}\times \\
    &\left[1+\nu e^{-\frac{\Delta t^{2}\sigma_{r}^{2}}{2\hbar^2}}\cos{\left(\frac{\Delta t}{\hbar}(E-E_{0})-\phi\right)}\right],
\end{split}
\end{equation}
where $P_\mathrm{max}$ and $\sigma_{E}$ are the peak power and the rms bandwidth of the envelope, $E_{c}$ is the central energy of the envelope, $E_0$ is the seed photon energy, $\sigma_{r}$ is the rms spectrometer resolution, $\nu$ is the contrast before convolution with the spectrometer resolution function, i.e. the contrast in the limit where  $\sigma_{r}\rightarrow 0$,  $\Delta t$ is the delay between neighbouring pulses, and $\phi$ is the phase of the interference pattern (\hyperref[Methods:spectrum]{Methods}).
Three example fits are shown in Fig.~\ref{fig_3}a and more are provided in the (\hyperref[supp:singleshots]{Supplementary Information}).
In the following, we discuss the mean values and the jitter of the fit parameters, which are shown as histograms in Fig.~\ref{fig_3}.
We exclude $\approx$25\% of the spectra, for which the fit does not converge.
The averaged spectral full-width-at-half-maximum bandwidth (FWHM) is \SI{1.8}{eV}, which is much larger than the measured bandwidth for standard self-seeding of \SI{0.19}{eV}~\cite{Nam:2021}.
This difference is consistent with a much shorter pulse duration for PHLUX.
Based on a FWHM resolution of \SI{0.26}{eV}~\cite{Kim:2025}, we estimate the contrast of the interference at $\nu$=$0.34\pm0.15$.
As we will show later, the pulse amplitude ratio between the head and tail pulses (Fig.~\ref{fig_1}a) can be estimated as $A_{3}/A_{1}\approx1/\sqrt{2}$.
The amplitude difference between pulses is caused by the beam tilt, therefore we assume $1/\sqrt{2}<A_{2}/A_{1}<1$.
Given the estimated amplitude ratio between the middle pulse with the head and tail pulses, we expect the contrast of two-pulse interferences to exceed the measured three-pulse contrast, implying that the two interferences act destructively.
The jitter of the contrast implies variations of the pulse amplitude ratios, caused by - among other effects - electron beam orbit jitter in the undulator section.
The orbit jitter is increased by the corrugated structure and, thus, is larger than for standard FEL configurations.

The separation of spectral lines shown in Figs.~\ref{fig_2} and \ref{fig_3} corresponds to a pulse delay of $8.1\pm1.0\,\mathrm{fs}$.
The fitted jitter of spectral line positions is \SI{0.11}{eV}, corresponding to \SI{1.2}{rad} of relative phase jitter between the middle and the sum of the other pulses.
At \SI{9.7}{keV} photon energy, this is equivalent to a relative carrier radiation arrival time jitter of \SI{0.1}{as} between neighbouring pulses, or a relative position jitter of the periodic electron bunching of \SI{30}{pm}.
The aforementioned PHLUX delay jitter of \SI{1.1}{fs} mainly concerns the envelope.
The central frequency jitter of the coherent output is \SI{0.3}{eV}, much larger than the crystal monochromator forward Bragg diffraction bandwidth of \SI{0.06}{eV}~\cite{Lindberg:2012,Min:2019,Nam:2021}, and also significantly larger than the spectral line position jitter of \SI{0.11}{eV}.
Effects that can cause a difference between the seed frequency and the central frequency of the self-seeded FEL are fluctuations in the monochromator crystal angle, as well as residual electron beam energy~\cite{Inoue:2019,Long:2025} and FEL pulse chirps.


\begin{figure}
  \centering
    \includegraphics[width=0.95\linewidth, trim=0cm 0cm 0cm 0cm, clip=true]{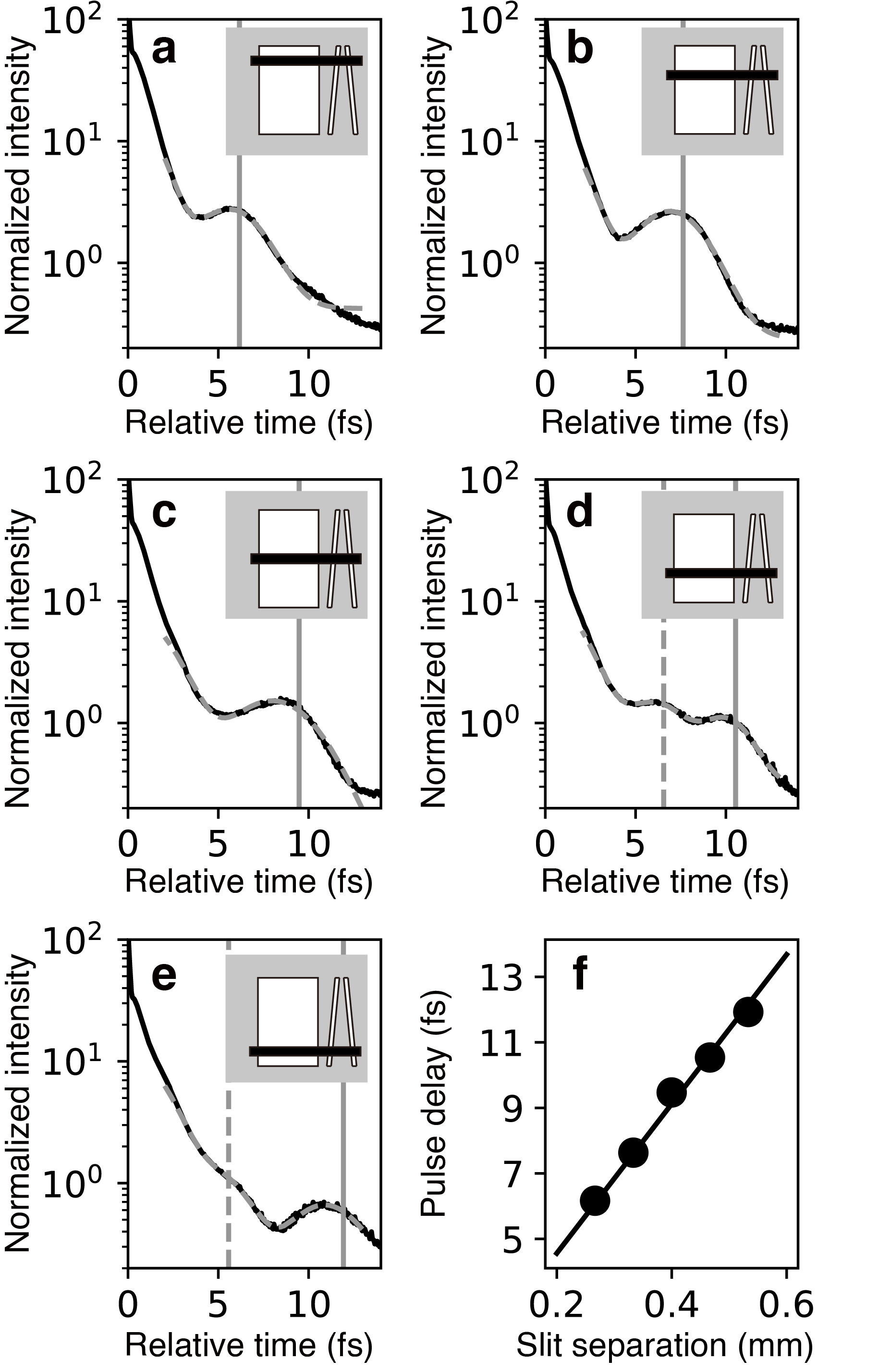}
    \caption{%
    \textbf{Pulse delay tunability}
    \textbf{a-e} Average of 1,000 single-shot PHLUX spectra measured for different positions of foil geometry~II (see insets).
    Solid lines are the averaged autocorrelation amplitudes, whereas dotted gray lines are fits to Eq.~(\ref{eq:Gauss}) for a-c with one peak visible, and Eq.~(\ref{eq:dGauss}) for d,e with two peaks.
    The vertical solid and dotted gray lines indicate the delay between the pulses generated by the first and last two slits, respectively.
    \textbf{f} Measured pulse delay as a function of slit center separation. The fitted slope, $\SI{23}{fs/mm}$, is related to electron beam properties (dispersion and energy chirp) at the slotted foil position.
    }
    \label{fig_4}
\end{figure}

\begin{figure}
  \centering
    \includegraphics[width=1.0\linewidth, trim=0cm 0cm 0cm 0cm, clip=true]{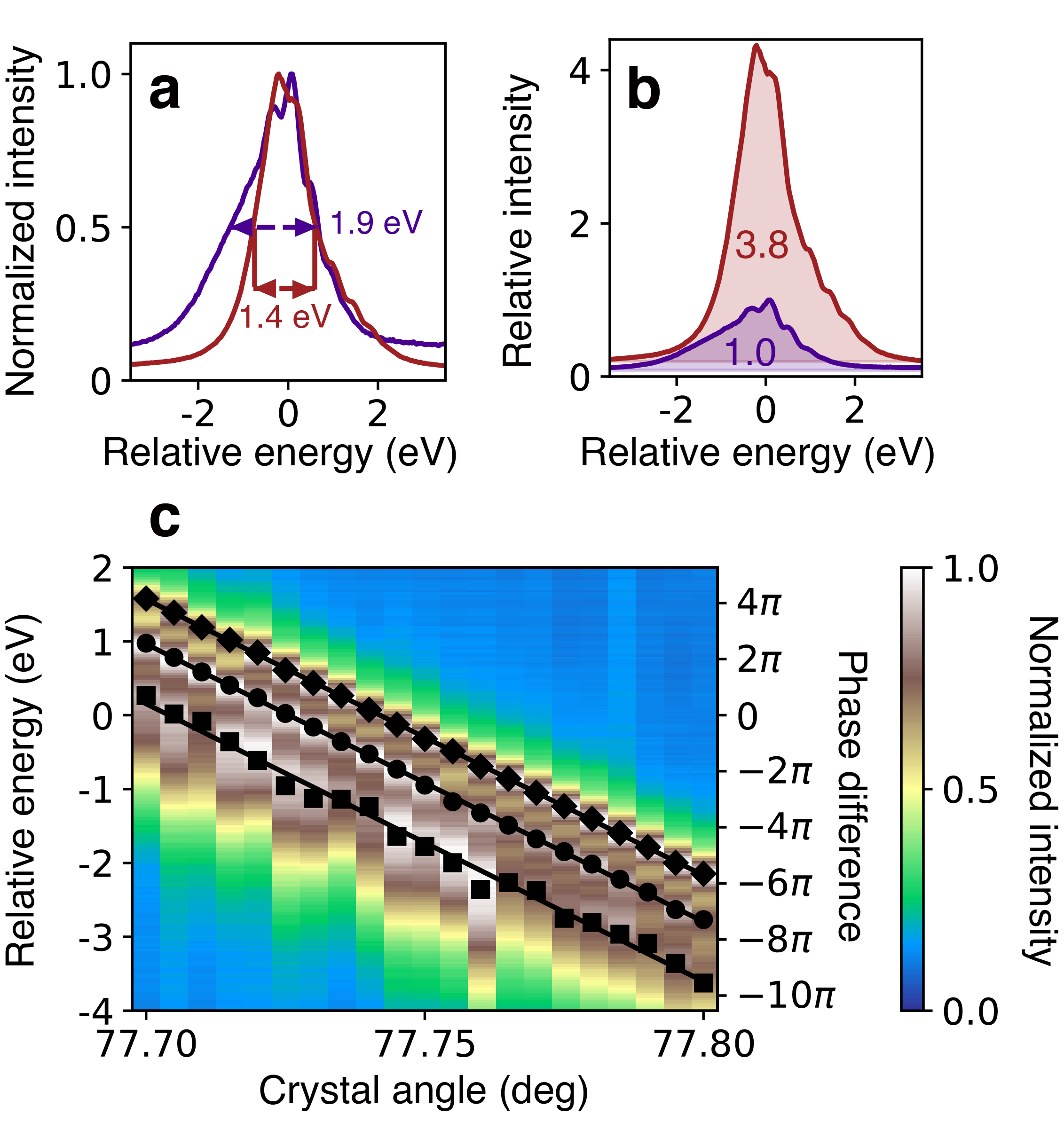}
    \caption{%
    \textbf{Pulse duration, amplitude and central frequency tunability}.
    \textbf{a} and \textbf{b} show the averaged spectrum using foil geometries~II (purple, \SI{0.1}{mm} head slit) and~III (red, \SI{0.2}{mm} head slit), respectively.
    The head slit centre separation is \SI{0.4}{mm} in both cases.
    The intensities are normalized either separately (a), or to the results from geometry~II~(b).
    \textbf{c} Averaged spectrum using foil geometry~I under variation of the monochromator crystal angle.
    We obtain three interference peak positions from each averaged spectrum by Gaussian fitting (black markers) and apply linear fits.
    }
    \label{fig_5}
\end{figure} 

We have characterized PHLUX performance in terms of delay, duration, amplitude, and central frequency, and now turn to its  tunability. 
 To show delay tunability, we use foil geometry~II (Fig.~\ref{fig_1}d), which contains two tilted slots and correlates the head pulse separation with the vertical foil position. 
 In particular, we scan the foil position by \SI{0.3}{mm} in five steps, acquire 1,000 spectra each, and fit the average autocorrelation amplitudes (Fig.~\ref{fig_4}) to obtain the pulse delay \mbox{(\hyperref[Methods:autocorrelation]{Methods})}.
Two of these measurements contain two peaks each (Fig.~\ref{fig_4}d,e), corresponding to distinguishable time separations of neighbouring pulses in the triplet.
Knowing the foil geometry and position, we determine which peak corresponds to the interference of the head pulses.
As expected, the delay is proportional to the slit center separation (Fig.~\ref{fig_4}f), ranging from 6.2 to \SI{11.9}{fs}.
To demonstrate an even shorter delay, we collect additional data with a foil geometry similar to geometry~I, but where the separation between the head slits is \SI{0.15}{mm} instead of \SI{0.2}{mm}, yielding a head pulse delay of \SI{4.5}{fs}.
From the autocorrelation curves that exhibit two peaks (Fig.~\ref{fig_4}d,e), we estimate that the head pulse carries about two times more pulse energy than the tail pulse, likely due to the progressively increasing beam tilt along the bunch that reduces the power of the tail pulse.
Using this information from the delay scan, we estimate a head pulse FWHM duration of \SI{3.8}{fs} (\hyperref[supp:foil]{Supplementary Information}).

We investigate also the impact of wider head slits by using foil geometry~III, where the width is twice compared to I and II.
Without the foil shaping resolution effect (\hyperref[supp:foil]{Supplementary Information}), the head pulse durations and energies would scale linearly with the slit width.
Considering this resolution effect, the FWHM head pulse durations are expected to increase only by factor 1.3 (from 3.8 to \SI{5.0}{fs}), consistent with the observation that the bandwidth of the coherent part of the spectrum decreases by factor 1.4 (from 1.9 to \SI{1.4}{eV}, Fig.~\ref{fig_5}a).
Moreover, the peak current of unspoiled electrons that pass the head slits is expected to increase by factor 1.6, and we find that the measured coherent pulse energy increases by a factor of $3.8$ (Fig.~\ref{fig_5}b), consistent with a larger FEL gain due to the higher current.

We also scan the phase of the interference pattern via the monochromator crystal angle which determines the seed photon energy.
The data are for geometry~I in a \SI{3}{eV} range, corresponding to a phase range of \mbox{$\approx10\pi$}.
Figure~\ref{fig_5}c shows the average of 100 spectra acquired for each of 21 steps and traces the three interference lines.
As expected, the interference pattern follows the moving spectrum envelope.

\section*{Discussion and outlook}

We have generated the first phase-locked hard X-ray pulses using a fresh bunch self-seeded FEL as well as a slotted emittance-spoiling foil, and measured the interference patterns generated by the coherent pulses.
Different slotted foil arrangements yield a tunable pulse delay ranging from 4.5 to \SI{11.9}{fs}. 
The average relative phase is freely varied via changing the monochromator setting.

Our demonstration of PHLUX is a milestone towards the development of hard X-ray arbitrary waveform generators (X-AWGs), with full amplitude and phase specification~\cite{Hastings:2019}.
Such X-AWGs will enable time-domain spectroscopies of the types used in the optical, THz and microwave regimes, as well as X-ray quantum optics, where coherent sequences of X-ray pulses will be used to control and read out quantum states. 
Applications will be to both basic physics and potentially also to the management of X-ray transparency and  damage for materials analysis and biomedicine.
Crucial will be not only an easily programmable and reproducible time structure for the X-ray waveforms, but also suitably large X-ray field amplitudes. 
The current experiment indicates the feasibility of reaching fields close to those needed for meaningful control of quantum states. 
In particular, based on a pulse duration of \SI{3.4}{fs}, a pulse energy of $\SI{20}{\uJ}$, and a focused beam spot of $340\times210$~nm$^2$ (available at the PAL-XFEL endstations~\cite{Kim:2025:2}), we estimate a peak electric field strength of about \SI{73}{GV\per\cm}.
To excite core electrons in atoms with an atomic number of around 25, this field strength corresponds to a Rabi cycle of $\approx\SI{3}{fs}$, close to the typical core-hole lifetime of $\approx\SI{1}{fs}$.

There is a clear programme of work which follows from our proof-of-concept experiments reported here. 
In particular, sources of instability in the seeded FEL need to be identified and remediated, and more advanced electron and photon beam diagnostics performed.
Also, the electron beam, prior to entering the first stage undulators, needs to be optimized to regulate the different pulse amplitudes and durations, including elimination of coherent second stage radiation from electrons passing the tail slit. 
This implies engineering both the beam phase space before the foil, as well as the foil itself, which may also enable sub-femtosecond pulse durations and time delays~\cite{Hartmann:2018}.
Eventually, we envision to shape the beam with a holographic laser heater~\cite{Cesar:2021}.
For exploitation and pulse characterization, higher resolution spectrometers will be employed.
Finally, we intend to generate PHLUX pulses also in the soft X-ray range, where the core-hole lifetimes are generally longer.

\section*{Methods}
\phantomsection
\noindent\textbf{Electron beam generation and shaping}\\
\label{Methods:beam}
A photoinjector generates bunches with \SI{180}{pC} charge. 
The bunches are accelerated in radio-frequency cavities and compressed in four-dipole chicanes.
The final beam energy and average current is \SI{8.54}{GeV} and \SI{3}{kA}, respectively~\cite{Kang:2017,Kang:2019}.

The emittance-spoiling Aluminium foil~\cite{Emma:2004} with \SI{3}{\um} thickness is installed in the last bunch compressor.
From the spectral analysis of measurements taken with foil geometry~I (Fig.~\ref{fig_2}c), we obtain a ratio between the PHLUX time separation and the slit center distance of \SI{27}{\unit{\fs\per\mm}}.
The pulse delay measurement as a function of slit center separation (Fig.~\ref{fig_4}f) was performed with a different machine and compression setup.
In this case we obtain a ratio of \SI{23}{\unit{\fs\per\mm}}.
We note that the slotted foil has a mechanical advantage by two orders of magnitude over an optical delaying stage that relies on differences in path length at the speed of light (corresponding to \SI{3335}{\unit{\fs\per\mm}}).
We estimate the rms time resolution of the slotted foil beam shaping as \SI{1.3}{fs} \mbox{(\hyperref[supp:foil]{Supplementary Information})}.
Under the circumstances discussed in this work, and confirmed by simulations shown in \hyperref[supp:gain_curve]{Supplementary Information}, an FEL pulse generated from a distribution of unspoiled electrons is of similar duration as the distribution itself.

The wakefield structure contains two parallel, individually movable Aluminium corrugated plates that are \SI{40}{mm} wide and \SI{1.4}{m} long.
The mechanical design and corrugation geometry are similar to those of the structure described in~\cite{Emma:2014}.
We calculate the wakefield effect using an analytical model~\cite{Bane:2016}, and estimate that distances between the beam and one of the plates in the range of 500 to \SI{700}{\um} are suitable to generate the beam tilts required for PHLUX.\\

\phantomsection
\noindent\textbf{Spectrum measurement and single-shot fits}\\
\label{Methods:spectrum}
A curved silicon crystal spectrometer records each FEL pulse spectrum.
Its FWHM resolution was previously determined as $0.26\pm0.04\,\mathrm{eV}$ at a photon energy of~\SI{9.7}{keV}~\cite{Nam:2021,Kim:2025}. 
This photon energy is a compromise between generating sufficient SASE seed pulse energy (favouring low photon energies) and a high spectrometer resolution (favouring high photon energies).

We now derive expressions for fitting the measured spectrum.
In time domain, the unspoiled electrons amplify the seed signal with a photon energy of $E_{0}$.
We assume the PHLUX pulses are transform-limited with identical rms pulse durations $\sigma_t$ and photon energy shifts $\delta E$ with respect to the seed photon energy $E_0$, originating from a residue linear beam energy chirp and undulator dispersion.
The resulting electric fields of the pulses are
\begin{equation}
\label{eq:1}
    \tilde{E}_{n}=A_{n}e^{i[\phi_{n}+\frac{\delta E}{\hbar}(t-t_{n})]}e^{-\frac{(t-t_{n})^{2}}{2\sigma_t^{2}}}e^{i\frac{E_{0}}{\hbar}t}, n=1,2,3.
\end{equation}
The term $A_{n}e^{i\phi_{n}}$ is proportional to the local FEL gain at $t_{n}$, which can vary along the electron beam.
The measured radiation power spectrum $P$ is the convolution of the true power spectrum $P_{0}$ and a normal distribution characterized by the spectrometer resolution ($2.355 \cdot\sigma_{r}=0.26\,\mathrm{eV}$), with
\begin{equation}
\label{eq:2}
    P_{0}=\varepsilon_{0}(\hat{E}_{1}+\hat{E}_{2}+\hat{E}_{3})(\hat{E}^{*}_{1}+\hat{E}^{*}_{2}+\hat{E}^{*}_{3}),
\end{equation}
where $\hat{E}_{1},\hat{E}_{2},\hat{E}_{3}$ are the spectral domain electric fields, and $\varepsilon_{0}$ is the vacuum permittivity.
As expected, Eq.~(\ref{eq:2}) shows that the total spectrum is the sum of individual pulse spectra and the interferences between all pulse pairs.
We add a constant term ($\mathrm{bg}$) in Eq.~(\ref{eq:2}) to account for the SASE signal, which within the range of interest ($\SI{9.7}{keV}\pm\SI{4}{eV}$) can be viewed as constant.
The spectrometer resolution has two effects on the spectra:
First, it broadens the total bandwidth from $\sigma_{E}=\hbar/\sqrt{2}\sigma_t$ to $\sigma'_{E}=\sqrt{\sigma_{E}^{2}+\sigma_{r}^{2}}$, which is negligible as $\sigma_{r}\ll \sigma_{E}$.
Second, it reduces the amplitude of the interference terms between pulses $n$ and $m$ ($P_{0nm}$, \mbox{$n,m=1,2,3$} with $n \neq m$) from $P_{0nm}=2\varepsilon_{0}A_{n}A_{m}$ to
\begin{equation}
\label{eq:3}
    P_{nm}=P_{0nm}e^{- \frac{(t_{n}-t_{m})^{2}}{2\hbar^{2}(\sigma_{E}^{-2}+\sigma_{r}^{-2})}}\approx P_{0nm}e^{-\frac{\sigma_{r}^{2}(t_{n}-t_{m})^{2}}{2\hbar^2}},
\end{equation}
which decays exponentially with the delay between pulses. 
This effect masks the interference between pulses 1 and 3 in the experiment.
When the time separation between neighbouring pulses is equal, we can further simplify the expression for $P$ to Eq.~(\ref{eq:fit}), where $P_\mathrm{max}=\varepsilon_{0}(A_{1}^{2}+A_{2}^{2}+A_{3}^{2})\sigma_t^{2}$ is the peak power and $E_{\mathrm{c}}=E_{0}+\delta E$ the central energy of the envelope. 
\begin{equation}
    \nu=\frac{2A_{2}\left|A_{1}e^{i(\phi_{1}-\phi_{2})}+A_{3}e^{i(\phi_{2}-\phi_{3})}\right|}{A_{1}^{2}+A_{2}^{2}+A_{3}^{2}}
\end{equation}
is the contrast (after deconvolution of the resolution), $\Delta t$ is the delay between neighbouring pulses, and $\phi$ is the complex angle of $A_{1}e^{i(\phi_{1}-\phi_{2})}+A_{3}e^{i(\phi_{2}-\phi_{3})}$.
The fit parameters of Eq.~(\ref{eq:fit}) are $\mathrm{bg}$, $P_\mathrm{max}$, $E_{c}$, $\sigma_{E}$, $\nu$, $\Delta t$, and $\phi$, and the total coherent pulse energy (as shown in Fig.~\ref{fig_2}b) is equal to $\sqrt{2\pi}P_\mathrm{max}\sigma_E$.


As demonstrated by the monochromator crystal angle scan (Fig.~\ref{fig_5}c), the phase of the interference, determining whether the spectral intensity from different pulses add up constructively or destructively at the reference photon energy $E_\mathrm{ref}$, is tunable via the seed photon energy.
If the time separation between neighbouring pulses is equal, the phase shift is given by $\Delta \phi = \Delta t (E_0-E_\mathrm{ref})/\hbar$~\cite{Reiche:2022}.\\

\phantomsection
\noindent\textbf{Time correlation function}
\label{Methods:correlation}

We calculate the first-order time correlation function $g_{1}$ of the PHLUX setup achieved with foil geometry~I, assuming it is a function of time delay.
The Wiener–Khinchin theorem equates the autocorrelation in time to the Fourier transform of power spectra, and lets us express $g_1$ in terms of $P_{0}$~\cite{Goodman:2015}:
\begin{equation}
    g_{1}(\Delta t)=\frac{|\langle\mathcal{F}\{P_{0}\}(\Delta t)\rangle|}{\sqrt{\langle|\mathcal{F}\{P_{0}\}(\Delta t)|^{2}\rangle}},
    \label{eq:g1}
\end{equation}
where angle brackets denote averaging over all spectra, and $\mathcal{F}$ denotes Fourier transformation.
Note that we can substitute $P$ for $P_0$ since the contribution of the spectrometer resolution cancels when calculating $g_1$.
The expression $|\langle\mathcal{F}\{P\}(\Delta t)\rangle|$ (orange line in Fig.~\ref{fig_2}d) is sensitive to both PHLUX delay and relative phase.
Conversely, the expression $\langle|\mathcal{F}\{P\}(\Delta t)|\rangle$ (blue line in Fig.~\ref{fig_2}d) is sensitive to the PHLUX delay but not the relative phase.
\\


\phantomsection
\noindent\textbf{PHLUX delay analysis}\\
\label{Methods:autocorrelation}
For the delay scans shown in Fig~\ref{fig_4}, we fit the averaged autocorrelation amplitudes with one or two Gaussian functions depending on the number of visible interference peaks, along with a Gaussian envelope and a constant background.
The fit functions are:
\begin{equation}
\label{eq:Gauss}
    y(t)=\left(A_{0}e^{-\frac{t^{2}}{2\sigma^{2}_{0}}}+A_{1}e^{-\frac{(t-t_{1})^{2}}{2\sigma^{2}_{1}}}\right)e^{-\frac{t^{2}}{2\sigma^{2}_{r}}}+\mathrm{bg},
\end{equation}
and
\begin{equation}
\label{eq:dGauss}
\begin{split}
    y(t)=&\left(A_{0}e^{-\frac{t^{2}}{2\sigma^{2}_{0}}}+A_{1}e^{-\frac{(t-t_{1})^{2}}{2\sigma^{2}_{1}}}+A_{2}e^{-\frac{(t-t_{2})^{2}}{2\sigma^{2}_{2}}}\right)\times \\
    & e^{-\frac{t^{2}}{2\sigma^{2}_{r}}}+\mathrm{bg},
\end{split}
\end{equation}
where $\sigma_{r}$ is the aforementioned spectrometer resolution, and $A_{0}$, $\sigma_{0}$, $A_{1}$, $t_{1}$, $\sigma_{1}$, $A_{2}$, $t_{2}$, $\sigma_{2}$, as well as bg are the fit parameters.
Parameters $A_{0}$ and $\sigma_0$ represent the main peak which results from the autocorrelation of individual pulses.
The Gaussian fit means $t_{1}$ and $t_{2}$ are the delays between the pulses, while the fit amplitudes $A_{1}$ and $A_{2}$ are proportional to the product of the interfering pulse amplitudes.

For cases of Fig.~\ref{fig_4}d,e with two visible interference peaks, one peak corresponds to the interference between the central and the head pulses, and the other to the interference between the central and tail pulses. 
Thus, one can estimate the averaged amplitude ratio between the head and tail pulses using the ratio of the fitted amplitudes, $A_{1}/A_{2}$.
According to these fits, the head pulse is about 1.7 times larger than the tail pulse for Fig.~\ref{fig_4}d, and 2.2 times larger for Fig.~\ref{fig_4}e.

\section*{Data availability}
Data supporting the plots of this paper and other findings of this study are available from the corresponding authors upon request.

\begin{acknowledgments}
The authors acknowledge the support from the Pohang Accelerator Laboratory operation group.
This work was supported by the National Research
Foundation of Korea (NRF) grants RS-2024-00455499, RS-2022-NR070321, and RS-2023-00237194, funded by the Korea government (MSIT).
Work at PSI was funded by the European Research Council under the European Union’s Horizon 2020 research and innovation program, within Grant Agreement 810451 (HERO).
\end{acknowledgments}

\section*{Author contributions}
G.A., H.-S.K., I.N., H.H., S.G., and S.R. conceived the project.
C.H.S, M.C., W.H., P.D., I.N., G.K., S-H.K., C-K.M., Y.J.S., and D.N. prepared the experiment.
C.H.S, M.C., W.H., P.D., C.-K.S, H.Y., S.K., K.-J.M., I.N., G.A., S.R., and E.P. performed the experiment.
W.H. and S.R. carried out the numerical simulations.
W.H., P.D., C.H.S. and M.C. analyzed the data and prepared the manuscript with input from all authors.

\section*{Competing interests}
The authors declare no competing interest.

\bibliography{export}

\vspace{\baselineskip}

\clearpage

\appendix
\section*{Supplementary Information}
\label{supp}

\subsection{Measured and simulated gain curve}
\label{supp:gain_curve}
\begin{figure}[h]
  \centering
    \includegraphics[width=\linewidth, trim=0cm 0cm 0cm 0cm, clip=true]{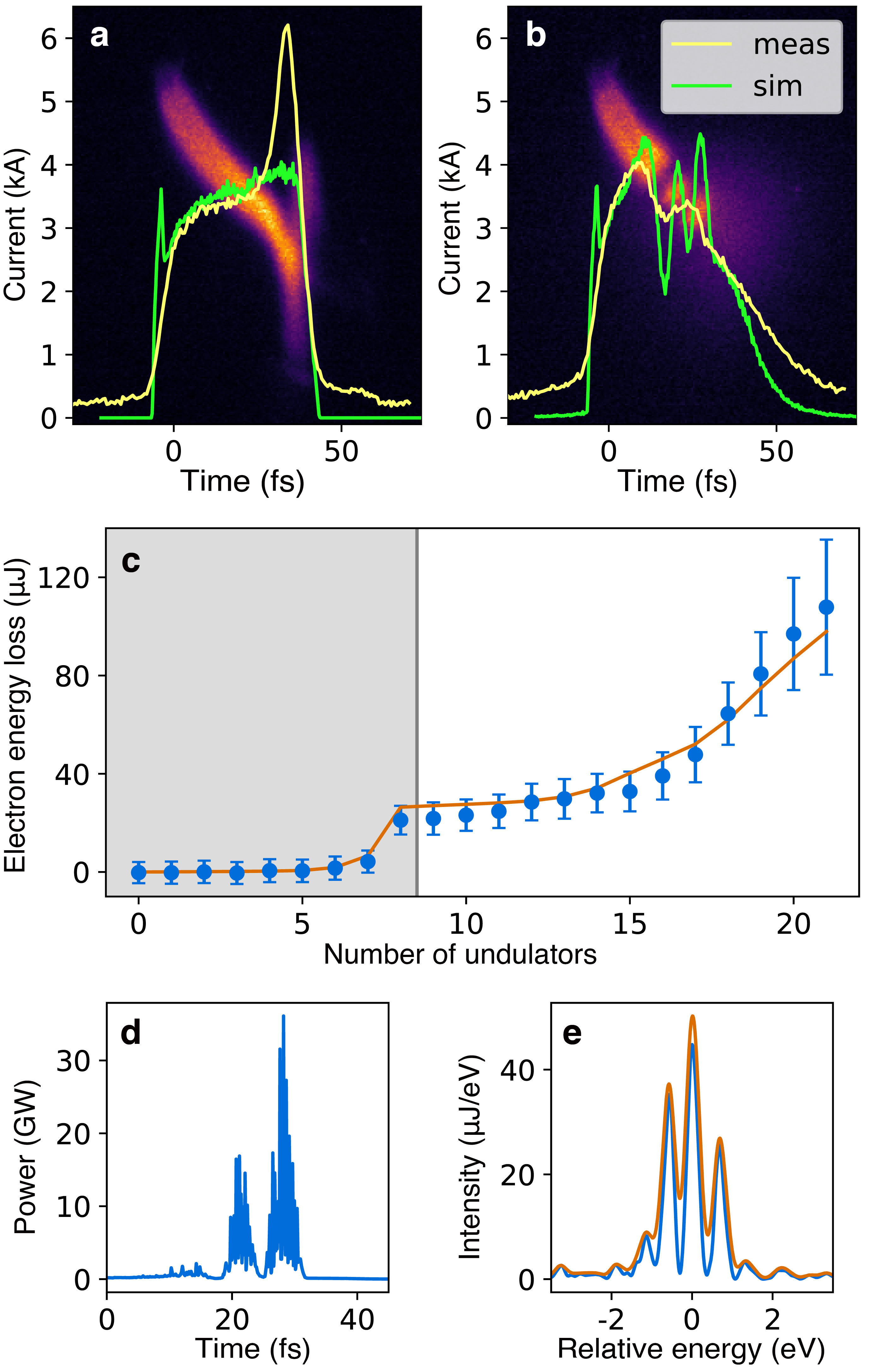}
    \caption{%
    \textbf{Measured FEL gain curves and PHLUX simulation}.
    \textbf{a} and \textbf{b} show the measured (yellow) beam longitudinal phase space without and with slotted foil (geometry~I in Fig.~\ref{fig_1}d), respectively.
    Green lines show the corresponding simulated current profiles.
    \textbf{c} Simulated (orange) and measured (blue) FEL pulse energy along the undulator beamline.
    The simulated results are scaled by a factor 1/2.3 for best agreement to the measurements.
    The first 8 undulators (gray area) are before the monochromator, separating the gain curve into two parts.
    \textbf{d}, \textbf{e} Blue lines depict the simulated output radiation in time (d) and frequency (e) domain.
    In time domain, the simulated pulse durations are $\approx\SI{3.6}{fs}$.
    The orange line visualizes the expected spectrum considering the spectrometer resolution.
    }
    \label{fig_supp_1}
\end{figure}

The gain curve describes the growth of the FEL pulse energy along the undulator beamline.
At PAL-XFEL, the gain curve is measured by recording the final electron beam energy while disabling the FEL process at various points along the undulator beamline~\cite{Loos:2011}, i.e. as a function of the number of undulator modules contributing to lasing.
We also simulate the gain curve using the \textit{elegant}~\cite{Borland:2000} and \textit{Genesis 1.3}~\cite{Reiche:1999} macroparticle codes for the electron beam shaping and FEL generation parts, respectively.
The simulation starts with an ideal beam before the last bunch compressor.
In particular, we assume a flat profile, with \SI{458}{A} beam current, \SI{105}{\micro m} bunch length, \SI{3.34}{GeV} central beam energy, \SI{70.1}{keV} rms energy spread, \SI{0.4}{\micro m} normalized transverse rms emittance, and a linear energy chirp of \SI{284}{keV/\micro m}.
We also consider that the electron beam is stretched at the middle of the bunch compressor.
We compare the \textit{elegant} simulation result with the measured beam current profile after the bunch compressor, and generally find a good agreement (Fig.~\ref{fig_supp_1}~a,b).
The peak at the bunch head results from non-linearities accumulated in the bunch compressors, which is not reflected in the simulation.
Nonetheless, with the slotted foil electrons at the horn do not contribute to lasing.
We use \textit{elegant}'s built-in slotted foil model to simulate foil geometry~I.
An analytical model~\cite{Bane:2016} provides us with the wake functions for the corrugated structure, where we assume a distance of \SI{600}{\um} between the beam and the nearby plate.
We halt the simulation before the structure, apply the wakefield effect to the particle distribution, rematch the transverse beam optics, and resume the simulations until the undulator entrance of the first stage.
The final macroparticle distribution is passed to \textit{Genesis~1.3}.
We adjust the undulator parameters of the second stage and the orbit alignment for best FEL performance.
The simulations overestimate the pulse energy by a factor of~$2.3$, as they use an ideal input beam and do not include imperfections that exist at the real facility, \textit{e.g.} optics mismatch, residual dispersion, and undulator misalignment.

The gain curve measurement reveals that the FEL does not reach deep saturation, indicating that the setup would benefit from a higher current or an increased number of undulators.
Driving the lasing process into saturation could make the output less susceptible to beam orbit jitter and, thereby, improve the output stability.

\subsection{Duration of the unspoiled electron distribution after the slotted foil}
\label{supp:foil}
The transverse divergence of particles hitting the slotted foil is greatly increased, preventing them from contributing to the FEL emission in the downstream undulator section.
Thus, the current profile of the unspoiled electrons is of interest.
The foil acts on electrons depending on their transverse position $x$, which is highly correlated to their longitudinal position $t$.
This correlation is a consequence of the simultaneous presence of linear dispersion $D=\Delta x/\Delta \delta$ (with relative momentum deviation $\delta$) caused by the bunch compressor dipoles, and linear energy chirp $k=\Delta \delta/\Delta t$ mostly caused by the upstream linear accelerators.
The combination of both effects yields $\Delta x/\Delta t=Dk$.
$\Delta t$ can be retrieved from spectral measurements as discussed in the main text.
Finally, with $D$ being a function of magnet parameters, and $\Delta x$ being a known foil property, we also determine $k$.

\begin{table}[t]
\centering
\begin{tabular}{|c|c|c|c|c|c|}
\hline $\sigma_{x}$ ($\mathrm{\mu m}$) & $\sigma_{y}$ ($\mathrm{\mu m}$) & $D$ (m) & $\sigma_{\delta}$ ($10^{-5}$) & $k$ ($\mathrm{ns}^{-1}$) & $b$ \\
\hline 56 & 33 & -0.21 & 17 & 207 & 0 or 1/9 \\
\hline
\end{tabular}
\caption{
    \textbf{Parameters to calculate the foil resolution.}
    Foil geometries I and III have zero slot tilt $b$, and geometry~II has $b=1/9$.
    }
\label{table:Table1}
\end{table}

\begin{figure}[h]
  \centering
  \includegraphics[width=\linewidth, trim=0cm 0cm 0cm 0cm, clip=true]{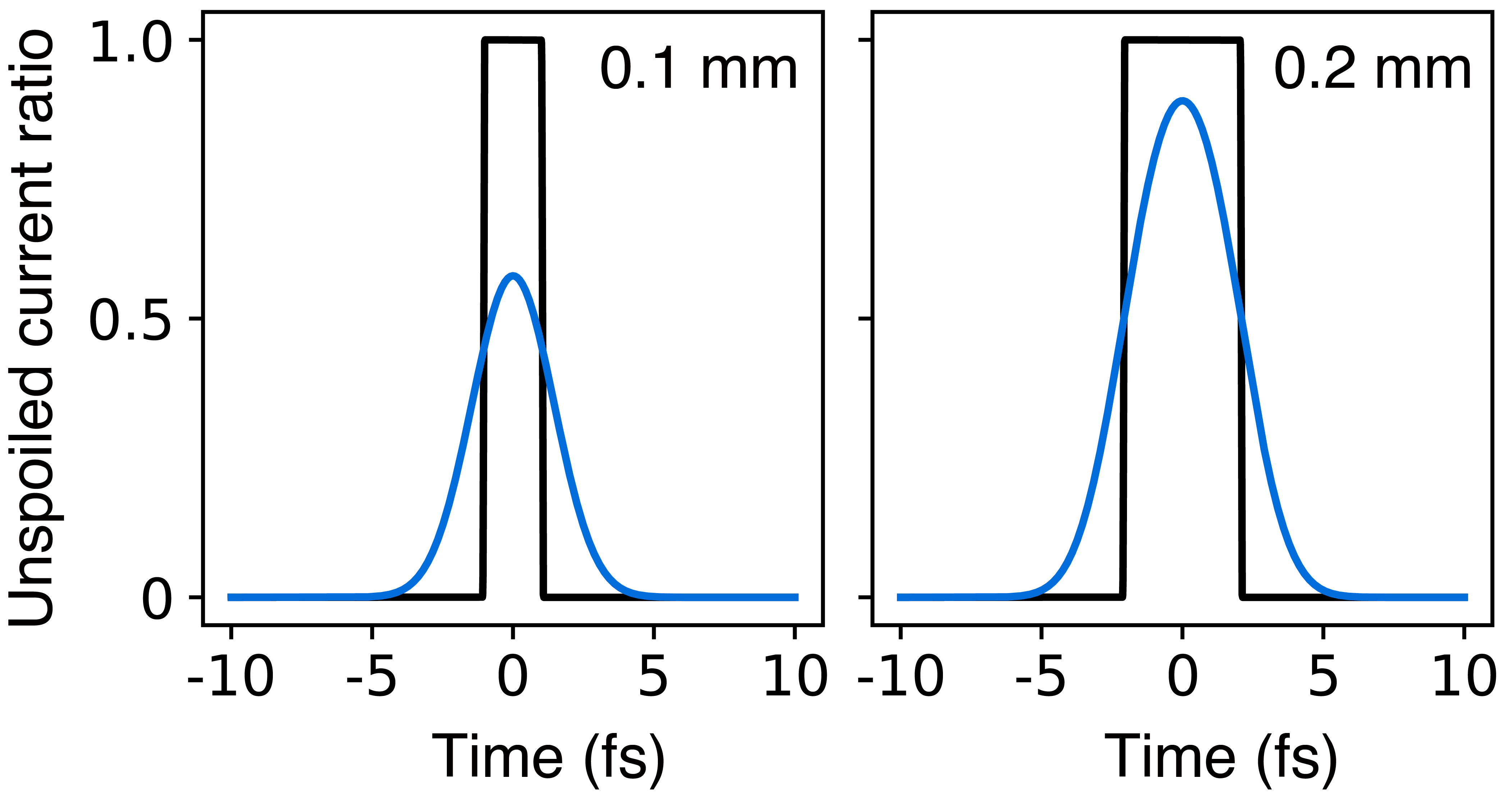}
    \caption{%
    \textbf{Distribution of the unspoiled electrons}.
    The blue lines show the calculated unspoiled current profile after the beam passing a \SI{0.1}{mm} (left) or \SI{0.2}{mm} wide slit (right) with straight edges.
    To calculate the resolution we use the parameters from Table~\ref{table:Table1}.
    The black lines show the case without considering the shaping resolution.
    }
    \label{fig_supp_2}
\end{figure}

The correlation between $x$ and $t$ is not absolute since the natural beam size and the uncorrelated energy spread also contribute to the $x$ coordinates.
Hence, the current profile of unspoiled electrons is not hard-edged.
The unspoiled electron current profile width after passing a foil with a single small slit was first calculated in~\cite{Emma:2004}.
Here we provide a more general expression for the rms shaping resolution $r_t$ which also accounts for slot edge tilts and uncorrelated energy spread:
\begin{equation}
r_t=\frac{\sqrt{\sigma_{x}^{2}+b^{2}\sigma_{y}^{2}+D^{2}\sigma_{\delta}^{2}}}{|Dk|},
\label{eq:foil_res}
\end{equation}
where $b$ reflects the angle of the foil edge which is defined as $y=bx$, $\sigma_{x}$ and $\sigma_{y}$ are the horizontal and vertical beam sizes at the foil position, and $\sigma_{\delta}$ is the normalized uncorrelated energy spread after the bunch compressor.
The unspoiled current profile is a convolution of a step function describing the slotted foil shape seen by the beam center and a normalized Gaussian function with rms width $r_t$.
Equation~(\ref{eq:foil_res}) for the parameters shown in Tab.~\ref{table:Table1} yields an $r_t$ of \SI{1.5}{fs} for both slot tilt angles used.
The dominant contribution to $r_t$ is the natural beam size $\sigma_x$.
The value for $k$ is chosen to match $|Dk|^{-1}=$~\SI{23}{\unit{\fs\per\mm}} and agrees to 10\% with the start-to-end electron beam simulations.

Figure~\ref{fig_supp_2} shows the unspoiled current profile for single slits of width \SI{0.1}{mm} (foil geometry~I) and \SI{0.2}{mm}~(geometry~III).
The calculated FWHM (rms) duration of the unspoiled electrons are \SI{3.8}{fs} (\SI{1.6}{fs}) for the \SI{0.1}{mm} wide slit and \SI{5.0}{fs} (\SI{2.0}{fs}) for the \SI{0.2}{mm} wide one.
The duration is resolution-limited in particular for the small slit, resulting in a reduced unspoiled peak current compared to the overall beam current.

\subsection{Single-shot spectra}
\label{supp:singleshots}
\begin{figure}[h]
  \centering
  \includegraphics[width=\linewidth, trim=0cm 0cm 0cm 0cm, clip=true]{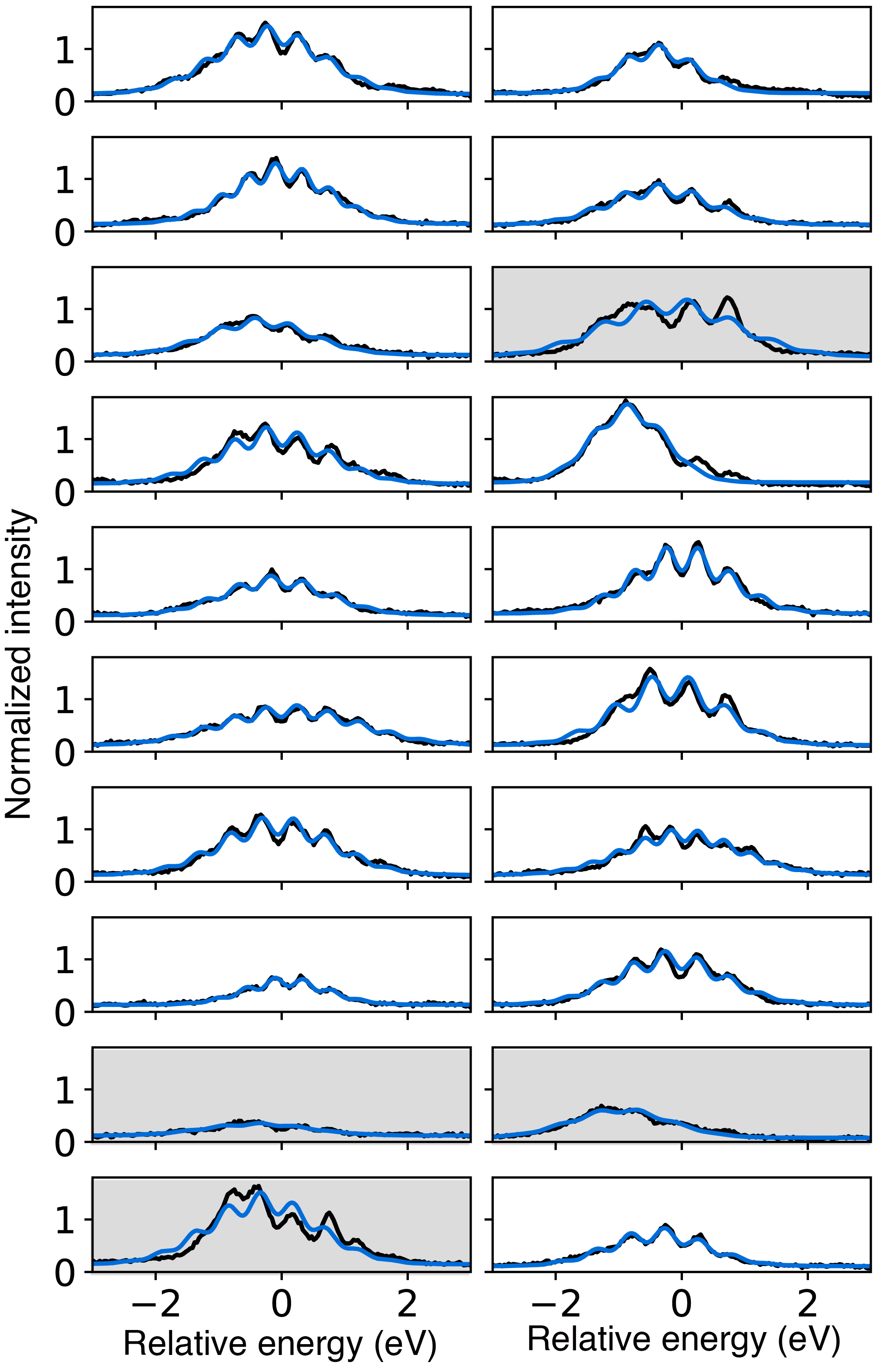}
    \caption{%
    \textbf{Single-shot spectra}.
    20 consecutive single-shot PHLUX spectra (black lines) taken using foil geometry~I.
    Blue lines are the corresponding fits (\hyperref[Methods:spectrum]{Methods}).
    For the shots with gray background we do not obtain satisfactory fits.
    }
    \label{fig_supp_5}
\end{figure}

Figure~\ref{fig_supp_5} shows 20 consecutive single-shot spectra out of 2,000 total, and their fits to Eq.~(\ref{eq:foil_res}), from the measurement shown in Figs.~\ref{fig_2},\ref{fig_3} of the main text.
We exclude 25\% of all spectra from further analysis, based on a too large fit residue, an implausible fitted pulse delay, or insufficient total spectrum amplitude (applies to four spectra in Fig.~\ref{fig_supp_5}).

\end{document}